\def\kl{\mathbf{\hat{k}}\mathbf\cdot\mathbf{\hat{l}}}
\def\km{\mathbf{\hat{k}}\mathbf\cdot\mathbf{\hat{m}}}
\def\kh{\mathbf{\hat{k}}\mathbf\cdot\mathbf{\hat{h}}}

\def\nk{n_{\rm b}}

\def\acap{\\ \nonumber \\}

\def\snf{\sin f}
\def\csf{\cos f}

\def\Pb{P_{\rm b}}

\def\rfr#1{Equation\,(\ref{#1})}
\def\rfrs#1#2{Equations\,(\ref{#1})--(\ref{#2})}
\def\Rfr#1{Equation\,(\ref{#1})}
\def\Rfrs#1#2{Equations\,(\ref{#1})--(\ref{#2})}

\def\dert#1#2{\frac{{{\textrm{d}}}{#1}}{{{\textrm{d}}}{#2}}}

\def\virg#1{``#1"}

\def\eqi{\begin{equation}}
\def\eqf{\end{equation}}
\def\eqia{\begin{eqnarray}}
\def\eqfa{\end{eqnarray}}

\def\rp#1#2{\frac{#1}{#2}}
\def\lb#1{\label{#1}}

\def\bds#1{\mathbf{#1}}

\def\cO{\cos\Omega}
\def\sO{\sin\Omega}

\def\cI{\cos I}
\def\sI{\sin I}

\def\ton#1{\left(#1\right)}
\def\qua#1{\left[#1\right]}
\def\grf#1{\left\{#1\right\}}
\def\ang#1{\left\langle #1\right\rangle}
\RequirePackage[2020-02-02]{latexrelease}
\documentclass{aastex631}
\usepackage{morefloats}
\usepackage[title]{appendix}
\usepackage{textcomp}
\usepackage{booktabs}
\usepackage{multirow}
\usepackage{rotating,tabularx}
\usepackage{float}
\usepackage{enumerate}
\usepackage{rotating}
\usepackage[polutonikogreek,english]{babel}
\usepackage{amsmath,starfont,textgreek,w-greek,wasysym}
\usepackage[flushleft]{threeparttable}
\usepackage{amsthm}
\usepackage{amscd}
\usepackage[mathlines]{lineno}
\usepackage{amssymb,dsfont}
\usepackage{graphicx,epsfig}
\usepackage{txfonts}
\usepackage{xr-hyper}
\usepackage{hyperref}
\allowdisplaybreaks

\makeatletter
 \DeclareRobustCommand\ref{%
    \@ifstar\@refstar\T@ref
  }%
  \DeclareRobustCommand\pageref{%
    \@ifstar\@pagerefstar\T@pageref
  }%
 \makeatother

\begin{document}

\title{Post-Newtonian Orbital Effects Induced by the Mass Quadrupole and Spin Octupole Moments of an Axisymmetric Body}

\shortauthors{L. Iorio}

\author[0000-0003-4949-2694]{Lorenzo Iorio}
\affiliation{Ministero dell' Istruzione e del Merito. Viale Unit\`{a} di Italia 68, I-70125, Bari (BA),
Italy}

\email{lorenzo.iorio@libero.it}

\begin{abstract}
The post-Newtonian orbital effects induced by the mass quadrupole and spin octupole moments of an isolated, oblate spheroid of constant density that is rigidly and uniformly rotating on the motion of a test particle are analytically worked out for an arbitrary orbital configuration and without any preferred orientation of the body's spin axis. The resulting expressions are specialized to the cases of (a) equatorial and (b) polar orbits. The opportunity offered by a hypothetical new spacecraft moving around Jupiter along a Juno-like highly elliptical, polar orbit to measure them is preliminarily studied. Although more difficult to be practically implemented, also the case of a less elliptical orbit is considered since it yields much larger figures for the relativistic effects of interest. The possibility of using the S stars orbiting the supermassive black hole in Sgr A$^\ast$  at the Galactic Center as probes to potentially constrain some parameters of the predicted extended mass distribution surrounding the hole by means of the aforementioned orbital effects  is briefly examined.
\end{abstract}


\section{Introduction}
The most known post-Newtonian (pN) orbital effects, which have been extensively tested so far in different  astronomical scenarios, are the gravitoelectric and gravitomagnetic precessions due to the mass monopole
and the spin-dipole moments of the central body which acts as source of the gravitational field, respectively.
The former is responsible for the well known, previously anomalous perihelion precession of Mercury in the Sun's field \citep{LeVer1859} of $42.98\,''$ per century (arcsec cty$^{-1}$) \citep{1986Natur.320...39N}, whose explanation by \citet{Ein15} was the first empirical success of his general theory of relativity (GTR).
Such a feature of motion was later repeatedly measured with radar measurements of Mercury itself \citep{1972PhRvL..28.1594S,1990grg..conf..313S}, of other inner planets \citep{1978AcAau...5...43A,1992mgm..conf..353A}, and of the asteroid Icarus \citep{1968PhRvL..20.1517S,1971AJ.....76..588S} as well.
In more recent times, stars in the Galactic Center \citep{2020A&A...636L...5G}, binary pulsars \citep{2006Sci...314...97K} and Earth's artificial satellites \citep{2010PhRvL.105w1103L,2014PhRvD..89h2002L} were also used.
The latter is the so-called Lense-Thirring effect \citep{1918PhyZ...19..156L,1984GReGr..16..711M}, which is induced by the angular momentum of the central spinning body. It is currently being experimentally investigated around the Earth with the geodetic satellites of the LAGEOS family; see, for example, \citet{2013CEJPh..11..531R,2011Ap&SS.331..351I,2020Univ....6..139L}, and references therein.
The gravitomagnetic precessions of the spins of some orbiting gyroscopes \citep{Pugh59,Schiff60} were detected in the field of the rotating Earth with the
dedicated Gravity Probe B  spaceborne mission \citep{Varenna74} to a $19\%$ accuracy level \citep{2011PhRvL.106v1101E,2015CQGra..32v4001E}, although the originally expected error was $\simeq 1\%$ \citep{2001LNP...562...52E}.

Less known features of motion are the pN gravitoelectric and gravitomagnetic effects associated with the asphericity of a central body induced by its mass quadrupole and spin octupole moments, respectively \citep{1988CeMec..42...81S,Sof89,1990CeMDA..47..205H,1991ercm.book.....B,1992CeMDA..53..293H,2014CQGra..31x5012P,2014PhRvD..89d4043W,2015IJMPD..2450067I,2015CeMDA.123....1M,2018RSOS....580640F,2018CeMDA.130...40S}.
To date, they have never been measured; a detailed study for a proposed satellite-based mission in the field of the Earth, dubbed Highly Elliptical Relativity Orbiter, can be found in \citet{2019Univ....5..165I}.

We will analytically work out, to the first pN (1pN) order, the net rates of change per orbit induced by the aforementioned departures from spherical symmetry of the source of the gravitational field without recurring to any simplifying assumption pertaining both the orbital configuration of the test particle and the orientation of the primary's spin axis in space. We will present our results in a form that should make their interpretation and use in specific situations at hand simple and direct. Then, we will look at some astronomical scenarios that could be favorable for their measurement. Such a task would enlarge the empirical basis of GTR extending it to phenomena, although just in the 1pN regime,  not yet tested. Conversely, they may provide further means to dynamically constrain, at least in principle, some key physical parameters of astronomical and astrophysical systems of interest.
We will look neither at the directly measurable quantities in real spacecraft-based missions such as range and range rates, nor at the actual data reduction procedure \citep{Moyer03}. Instead, our goal is just to gain meaningful insights about the potential offered by the considered scenario(s) by performing a preliminary sensitivity analysis. To this aim, we will use the usual osculating Keplerian orbital elements \citep{Sof89,1991ercm.book.....B,2011rcms.book.....K,SoffelHan19}, which are easy to visualize in view of their clear geometric meaning.

The paper is organized as follows.
In Section\,\ref{compu}, the calculation scheme adopted to work out the long-term orbital features of motion under investigation is reviewed.
The case of the pN gravitoelectric static field of a massive, oblate body is treated in Section\,\ref{AJ2}, while the consequences of the gravitomagnetic spin octupole moment of a spinning primary is the subject of Section\,\ref{Soct}. Section\,\ref{orbi} deals with some particular orbital configurations; in Section\,\ref{equa}, the scenario where the satellite's orbital plane and the body's equatorial plane coincide (equatorial orbit) is considered, while Section\,\ref{pola} investigates the case where the orbital plane contains the body's spin axis (polar orbit). The results of Section\,\ref{pola} are applied in Section\,\ref{jup} to a Jovian scenario characterized by a putative spacecraft orbiting the fifth planet of our solar system along a highly elliptical polar orbit. A possible application of our results to the highly elliptical stellar motions in Sgr A$^\ast$ at the center of the Galaxy  is outlined in Section\,\ref{SgrA}. Section\,\ref{fine} summarizes our findings and offers concluding remarks.
Appendix\,\ref{appenA} contains a list of the symbols and definitions used throughout the paper.
\section{Computational outline}\lb{compu}
The long-term orbital effects of interest are calculated by averaging over one satellite's orbital period the right-hand-sides of the equations for the variations of the Keplerian osculating elements in the Euler--Gauss form \citep{Sof89,1991ercm.book.....B,2011rcms.book.....K,SoffelHan19}
\begin{align}
\dert{a}{t} \lb{dadt} &= \rp{2}{\nk\,\sqrt{1-e^2}}\,\qua{e\,A_R\,\sin f + \ton{\rp{p}{r}}\,A_T},\\ \nonumber \\
\dert e t \lb{dedt} & = \rp{\sqrt{1-e^2}}{\nk\,a}\,\grf{A_R\,\snf + A_T\,\qua{\csf + \rp{1}{e}\,\ton{1-\rp{r}{a}} }}, \\ \nonumber\\
\dert I t & = \rp{1}{\nk\,a\,\sqrt{1-e^2}}\,A_N\,\ton{\rp{r}{a}}\,\cos u, \\ \nonumber\\
\dert \Omega t & = \rp{1}{\nk\,a\,\sin I\,\sqrt{1-e^2}}\,A_N\,\ton{\rp{r}{a}}\,\sin u, \\ \nonumber\\
\dert \omega t \lb{dodt} & = \rp{\sqrt{1-e^2}}{\nk\,a\,e}\,\qua{-A_R\,\csf + A_T\,\ton{1 + \rp{r}{p}}\,\snf} - \cos I\,\dert\Omega t, \\ \nonumber \\
\dert\eta{t} \lb{detadt} &= -\rp{2}{\nk\,a}\,A_R\,\ton{\rp{r}{a}} - \rp{\ton{1-e^2}}{\nk\,a\,e}\,\qua{-A_R\,\cos f +A_T\,\ton{1+\rp{r}{p}}\,\sin f},
\end{align}
evaluated onto the Keplerian ellipse
\eqi
r = \rp{p}{1 + e~\cos f},
\eqf
by means of
\eqi
\dert t f = \rp{r^2}{\sqrt{\mu~p}}.
\eqf
It turns out to be computationally convenient to express the position and velocity vectors as
%
%
%
\begin{align}
\bds r \lb{rvec}& = r\ton{\bds{\hat{l}} \cos u + \bds{\hat{m}} \sin u}, \\ \nonumber \\
\bds v \lb{vvec}& = \sqrt{\rp{\mu}{p}}\qua{-\bds{\hat{l}}\ton{e \sin\omega + \sin u}  + \bds{\hat{m}}\ton{e \cos\omega + \cos u} }.
\end{align}
In the following, for the sake of simplicity, the angular brackets $\ang{\mathrm{d}\kappa/\mathrm{d}t}$ denoting the average over one orbital period of the rate of change of any one of the orbital elements $\kappa$  will be omitted.
\section{The 1pN acceleration induced by the oblateness of the central body}\lb{AJ2}
The pN gravitoelectric acceleration ${\mathbf{A}}_\mathrm{pN}^\mathrm{obl}$ experienced by a test particle orbiting an extended, oblate body is \citep{2014PhRvD..89d4043W}
\begin{equation}
{\mathbf{A}}_\mathrm{pN}^\mathrm{obl} = {\mathbf{A}}^\mathrm{I} + {\mathbf{A}}^\mathrm{II} + {\mathbf{A}}^\mathrm{III},\lb{AJ2pN}
\end{equation}
where
\begin{align}
{\mathbf{A}}^\mathrm{I} \label{A1} & = \frac{3~\mu~J_2~R_\mathrm{e}^2}{2~c^2~r^4}~\left(v^2 - \frac{4~\mu}{r}\right)\left[\left(5~\xi^2 - 1\right)\mathbf{\hat{r}} -2~\xi~\mathbf{\hat{k}}\right], \\ \nonumber \\
{\mathbf{A}}^\mathrm{II} \label{A2} & = -\frac{6~\mu~J_2~R_\mathrm{e}^2}{c^2~r^4}~\left[\left(5~\xi^2 - 1\right)~v_r
- 2~\xi~\lambda\right]~\mathbf{v}, \\ \nonumber \\
{\mathbf{A}}^\mathrm{III} \label{A3} & = -\frac{2~\mu^2~J_2~R_\mathrm{e}^2}{c^2~r^5}~\left(3~\xi^2 -1\right)~\mathbf{\hat{r}}.
\end{align}
%
%
${\bds A}_\mathrm{pN}^\mathrm{obl}$ was derived earlier by \citet{1988CeMec..42...81S,Sof89,1991ercm.book.....B,1992CeMDA..53..293H} in a reference frame whose  $z$-axis is aligned with $\bds{\hat{k}}$.

By following the computational scheme outlined in Section\,\ref{compu}, one obtains the following expressions for the long-term rates of change of the Keplerian osculating elements induced by the sum of \rfrs{A1}{A3}.
\begin{align}
\dert a t \lb{dadtBS}& = -\rp{9~e^2~\ton{6 + e^2}~\nk~J_2~\mu~R_\mathrm{e}^2~\ton{\textcolor{black}{\widehat{T}_3}~\sin 2\omega -2~\textcolor{black}{\widehat{T}_6}~\cos 2\omega}}{
8~c^2~a^2~\ton{1 - e^2}^4}, \\ \nonumber \\
\dert e t \lb{dedtBS}& = -\rp{21~e~\ton{2 + e^2}~\nk~J_2~\mu~R_\mathrm{e}^2~\ton{\textcolor{black}{\widehat{T}_3}~\sin 2\omega -2~\textcolor{black}{\widehat{T}_6}~\cos 2\omega}}{
16~c^2~a^3~\ton{1 - e^2}^3}, \\ \nonumber\\
\dert I t \lb{dIdtBS}& = \rp{3 ~\nk~J_2~\mu~R_\mathrm{e}^2~\qua{\textcolor{black}{\widehat{T}_4}~\ton{6 + e^2~\cos 2\omega} + e^2~\textcolor{black}{\widehat{T}_5}~\sin 2\omega}}{
4~c^2~a^3~\ton{1 - e^2}^3},\\ \nonumber \\
\dert \Omega t \lb{dOdtBS}& = -\rp{3~\nk~J_2~\mu~R_\mathrm{e}^2~\csc I~\qua{- e^2~\textcolor{black}{\widehat{T}_4}~\sin 2\omega + \textcolor{black}{\widehat{T}_5}~\ton{-6 + e^2~\cos 2\omega}}}{
4~c^2~a^3~\ton{1 - e^2}^3},\\ \nonumber \\
\dert \omega t \nonumber \lb{dodtBS}& = -\rp{3~\nk~J_2~\mu~R_\mathrm{e}^2 }{
16~c^2~a^3~\ton{1 - e^2}^3}\grf{
\ton{-8 + 3~e^2} \ton{-2~\textcolor{black}{\widehat{T}_1} + 3~\textcolor{black}{\widehat{T}_2}} + 14~\textcolor{black}{\widehat{T}_3}~\cos 2\omega + \right.\\ \nonumber \\
& +\left. 4~\qua{e^2~\textcolor{black}{\widehat{T}_4}~\cot I~\sin 2\omega +  \textcolor{black}{\widehat{T}_5}~\cot I~\ton{6 - e^2~\cos 2\omega} + 7~\textcolor{black}{\widehat{T}_6}~\sin 2\omega}
},\\ \nonumber \\
\dert \eta t  \lb{detdtBS}& = \rp{\nk~J_2~\mu~R_\mathrm{e}^2 }{
16~c^2~a^3~\ton{1 - e^2}^{5/2}}~\qua{
\ton{80 + 73e^2}~\ton{-2~\textcolor{black}{\widehat{T}_1} + 3~\textcolor{black}{\widehat{T}_2}} + 42~\ton{1 + 2~e^2}~\ton{\textcolor{black}{\widehat{T}_3}~\cos 2\omega + 2~\textcolor{black}{\widehat{T}_6}~\sin 2 \omega}
}.
\end{align}
\textcolor{black}{The coefficients $\widehat{T}_j,\,j = 1,\,2,\ldots 6$ are explicitly displayed in Appendix\,\ref{appenB}.}
It should be remarked that \rfrs{dadtBS}{detdtBS} are valid for any orbital configuration and for an arbitrary orientation of the body's symmetry axis in space. They were calculated in \citet{1988CeMec..42...81S,1991ercm.book.....B,1992CeMDA..53..293H} by orienting $\bds{\hat{k}}$ along the  $z$ axis of the reference frame chosen.
\citet{2015IJMPD..2450067I} worked out the pN oblateness-driven net shifts per orbit $\Delta\kappa$ of $\kappa = \grf{p,\,e,\,I,\,\Omega,\,\omega}$ for an arbitrary orientation of $\bds{\hat{k}}$, but the resulting expressions are much more cumbersome than \rfrs{dadtBS}{detdtBS} and a comparison with them is difficult.

The mixed effects due to the simultaneous presence of the 1pN gravitoelectric acceleration due to the mass monopole of the central body and the Newtonian acceleration induced by $J_2$ in the equations of motion \citep{1990CeMDA..47..205H,1992CeMDA..53..293H,2014PhRvD..89d4043W,2015IJMPD..2450067I} will not be calculated here\textcolor{black}{; they can be found, worked out in their full generality, in \citet{IorioGRG23}.} From an empirical point of view, it can be expected that they would affect post-fit residuals just with tiny mismodelled signatures since both the aforementioned accelerations are routinely modeled in any software used to reduce astronomical observations. \textcolor{black}{In principle, also orbital variations of order $\mathcal{O}\ton{J_2^2/c^2}$, arising from the mixed action of \rfr{AJ2pN} and the Newtonian oblateness-driven acceleration, should occur. Nonetheless, they are completely negligible since they would bring about in \rfrs{dadtBS}{detdtBS} the scaling factor $J_2\ton{R_\mathrm{e}/a}^2$.}
\section{The gravitomagnetic acceleration induced by the spin octupole moment of a rotating body}\lb{Soct}
To the 1pN order, the gravitomagnetic  Panhans-Soffel (PS) spin octupole\footnote{The spin-dipole moment in \citet{2014CQGra..31x5012P} yields the usual Lense-Thirring acceleration \citep{iers10}. For other studies on relativistic multipoles, see, e.g., \citet{2015CeMDA.123....1M,2018CeMDA.130...40S,2018RSOS....580640F}.} acceleration ${\mathbf{A}}^\mathrm{oct}_\mathrm{PS}$  experienced by a test particle orbiting an oblate spheroid of constant density that is rigidly and uniformly rotating is \citep{2014CQGra..31x5012P}
\begin{equation}
{\mathbf{A}}^\mathrm{oct}_\mathrm{PS} = \frac{\mathbf{v}}{c^2}\mathbf{\times}{\mathbf{B}}^\mathrm{oct},\label{eq1}
\end{equation}
where the gravitomagnetic field ${\mathbf{B}}^\mathrm{oct}$ can be calculated as \cite{2014CQGra..31x5012P}
\begin{equation}
{\mathbf{B}}^\mathrm{oct} = -{\mathbf{\nabla}} \phi^\mathrm{oct},
\end{equation}
with the gravitomagnetic octupolar potential $\phi^\mathrm{oct}$ given by \cite{2014CQGra..31x5012P}
\begin{equation}
\phi^\mathrm{oct} = \frac{6~G~S~R_\mathrm{e}^2~\varepsilon^2}{7~r^4}~\mathcal{P}_3\left(\xi\right).\label{eq3}
\end{equation}
From \rfrs{eq1}{eq3}, the spin octupole PS acceleration can finally be cast into the form
\begin{equation}
{\mathbf{A}}^\mathrm{oct}_\mathrm{PS} = \frac{3~G~S~R^2_\mathrm{e}~\varepsilon^2}{7~c^2~r^5}~\mathbf{v}\mathbf{\times}\left[5~\xi\left(7~\xi^2 - 3\right)~\mathbf{\hat{r}} + 3~\left(1 - 5~\xi^2\right)\mathbf{\hat{k}}\right],\label{eq4}
\end{equation}

According to the computational scheme outlined in Section\,\ref{compu}, the long-term rates of change of the osculating Keplerian orbital elements induced by \rfr{eq4} turn out to be
\begin{align}
\dert a t \lb{dadtPS}& = 0, \\ \nonumber\\
\dert e t \lb{dedtPS}& = \rp{45~e~G~S~R_\mathrm{e}^2~\varepsilon^2~\ton{\kh}~\ton{~\textcolor{black}{\widehat{T}_3}~\sin 2\omega - 2~\textcolor{black}{\widehat{T}_6}~\cos 2\omega}}
{28~c^2~a^5~\ton{1-e^2}^{5/2}}, \\ \nonumber\\
\dert I t \nonumber \lb{dIdtPS}& = -\rp{9~G~S~R_\mathrm{e}^2~\varepsilon^2}{56~c^2~a^5~\ton{1-e^2}^{7/2}}\ton{
2 \ton{2 + 3~e^2}~\ton{\kl}~\ton{-4~\textcolor{black}{\widehat{T}_1} + 5~\textcolor{black}{\widehat{T}_2}} +\right.\\ \nonumber \\
&+\left.  5~e^2~\grf{\ton{\kl}~\qua{-2~\textcolor{black}{\widehat{T}_1} + 3 \ton{\kl}^2 + \ton{\km}^2}~\cos 2\omega + 2~\ton{\km}~\qua{-\textcolor{black}{\widehat{T}_1} + 2 \ton{\kl}^2 + \ton{\km}^2}~\sin 2 \omega}
},\\ \nonumber \\
\dert\Omega t \nonumber \lb{dOdtPS}& = -\rp{9~G~S~R_\mathrm{e}^2~\varepsilon^2~\csc I}{56~c^2~a^5~\ton{1-e^2}^{7/2}}\ton{
2 \ton{2 + 3~e^2}~\ton{\km}~\ton{-4~\textcolor{black}{\widehat{T}_1} + 5~\textcolor{black}{\widehat{T}_2}} +\right.\\ \nonumber \\
&+\left.  5~e^2~\grf{-\ton{\km}~\qua{-2~\textcolor{black}{\widehat{T}_1} + \ton{\kl}^2 + 3 \ton{\km}^2}~\cos 2\omega + 2~\ton{\kl}~\qua{-\textcolor{black}{\widehat{T}_1} + \ton{\kl}^2 + 2 \ton{\km}^2}~\sin 2\omega}
},\\ \nonumber\\
\dert \omega t \nonumber \lb{dodtPS}& = \rp{9~G~S~R_\mathrm{e}^2~\varepsilon^2 }{56~c^2~a^5~\ton{1-e^2}^{7/2}}\ton{
4 \ton{3 + 2~e^2}~\ton{\kh}~\ton{-2~\textcolor{black}{\widehat{T}_1} + 5~\textcolor{black}{\widehat{T}_2}} +\right.\\ \nonumber \\
\nonumber &+\left. 2 \ton{2 + 3~e^2}\ton{\km}~\ton{-4~\textcolor{black}{\widehat{T}_1} + 5~\textcolor{black}{\widehat{T}_2}~\cot I}  + 5~\grf{2~\ton{1 + 2~e^2}~\ton{\kh}~\textcolor{black}{\widehat{T}_3} - \right.\right.\\ \nonumber \\
\nonumber &-\left.\left.    e^2~\ton{\km}~\qua{-2  +\ton{\kl}^2 + 3\ton{\km}^2}~\cot I }~\cos 2\omega +
 10~\ton{\kl}~\grf{2~\ton{1 + 2~e^2}~\textcolor{black}{\widehat{T}_5} + \right.\right.\acap
&\left.\left. + e^2~\qua{-1 + \ton{\kl}^2 + 2 \ton{\km}^2}~\cot I }~\sin 2\omega},\\ \nonumber \\
\dert\eta t \lb{detdtPS}&= -\rp{9~G~S~R_\mathrm{e}^2~\varepsilon^2~\ton{\kh}}{28~c^2~a^5~\ton{1-e^2}^2}~\grf{
5~\textcolor{black}{\widehat{T}_3}~\cos 2\omega +  2 \qua{-2~\textcolor{black}{\widehat{T}_1} + 5~\ton{\textcolor{black}{\widehat{T}_2} + \textcolor{black}{\widehat{T}_6}~\sin 2\omega}}
}.
\end{align}
It must be remarked that \rfrs{dadtPS}{detdtPS} retain their validity for any orientation of $\bds{\hat{k}}$ in space, and for arbitrary orbital configurations.
The gravitomagnetic spin octupole orbital precessions were explicitly calculated by \citet{2019MNRAS.484.4811I,2019MNRAS.485.4090I} in the special case of $\bds{\hat{k}}$ aligned with the  $z$ axis of the reference frame adopted. In fact, also general formulas for them, valid for arbitrary orbital geometries and orientations of $\bds{\hat{k}}$ in space, were obtained by \citet{2019MNRAS.484.4811I,2019MNRAS.485.4090I}; nonetheless, they  are much more cumbersome and less compact than \rfrs{dadtPS}{detdtPS}, so that a straightforward comparison is difficult. \textcolor{black}{The interplay of the Newtonian acceleration due to $J_2$ and the pN one of \rfr{eq4} would induce, in principle, mixed effects of order $\mathcal{O}\ton{S\,\varepsilon^2\,J_2/c^2}$. Nonetheless, also in this case, they would be negligible because of the multiplicative factor $J_2\,\ton{R_\mathrm{e}/a}^2$ by which \rfrs{dadtPS}{detdtPS} would be scaled down.}
\section{Some special orbital configurations}\lb{orbi}
Here, two particular orbital configurations are considered: (a) equatorial (Section\,\ref{equa}) and (b) polar (Section\,\ref{pola}) orbits.

By parameterizing $\bds{\hat{k}}$ in terms of the polar angles\footnote{If referred to the International Celestial Reference Frame (ICRF), $\alpha$ and $\delta$ are the R.A. and decl. of the primary's north pole of rotation, respectively \citep{2007CeMDA..98..155S}.} $\alpha,\,\delta$ as in Appendix\,\ref{appenA},
%
%
%
one has
\begin{align}
\kl \lb{kl}& = \cos\delta\,\cos\ton{\alpha - \Omega}, \\ \nonumber \\
\km \lb{km}& = \sin I\,\sin\delta + \cos I\,\cos\delta\,\sin\ton{\alpha - \Omega}, \\ \nonumber \\
\kh \lb{kh}& = \cos I\,\sin\delta - \sin I\,\cos\delta\,\sin\ton{\alpha - \Omega}.
\end{align}
\subsection{Equatorial orbit}\lb{equa}
Let us assume that the satellite's orbital plane coincides with the body's equatorial plane, irrespectively of the orientation of the latter in the adopted reference frame, i.e., for generic values of $\alpha,\,\delta$: it is
\begin{align}
\kl \lb{c1}& = \km =0, \\ \nonumber \\
\kh \lb{c2}& =1.
\end{align}
According to \rfrs{kl}{kh}, the conditions of \rfrs{c1}{c2} are fulfilled for
\begin{align}
I \lb{I1}& = -\delta +\rp{\pi}{2}, \\ \nonumber \\
\Omega \lb{O1}& = \alpha + \rp{\pi}{2}.
\end{align}

Then, \rfrs{dadtBS}{detdtBS} reduce to
\begin{align}
\dert a t \lb{dadtBSeq}& = 0, \\ \nonumber \\
\dert e t \lb{dedtBSeq}& = 0, \\ \nonumber\\
\dert I t \lb{dIdtBSeq}& = 0,\\ \nonumber \\
\dert \Omega t \lb{dOdtBSeq}& = 0,\\ \nonumber \\
\dert \omega t \lb{dodtBSeq}& = -\rp{3~\nk~J_2~\mu~R_\mathrm{e}^2~\ton{8 - 3~e^2} }{
8~c^2~a^3~\ton{1 - e^2}^3},\\ \nonumber \\
\dert \eta t \lb{detdtBSeq}& = -\rp{\nk~J_2~\mu~R_\mathrm{e}^2~\ton{80 + 73~e^2} }{
8~c^2~a^3~\ton{1 - e^2}^{5/2}},
\end{align}
while \rfrs{dadtPS}{detdtPS} can be written as
\begin{align}
\dert a t \lb{dadtPSeq}& = 0, \\ \nonumber \\
\dert e t \lb{dedtPSeq}& = 0, \\ \nonumber\\
\dert I t \lb{dIdtPSeq}& = 0,\\ \nonumber \\
\dert \Omega t \lb{dOdtPSeq}& = 0,\\ \nonumber \\
\dert \omega t \lb{dodtPSeq}& = -\rp{9~G~S~R^2_\mathrm{e}~\varepsilon^2~\ton{3 + 2~e^2}}{7~c^2~a^5~\ton{1-e^2}^{7/2}},\\ \nonumber \\
\dert \eta t \lb{detdtPSeq}& = \rp{9~G~S~R^2_\mathrm{e}~\varepsilon^2}{7~c^2~a^5~\ton{1-e^2}^{2}}.
\end{align}
Note that \rfrs{dodtBSeq}{detdtBSeq} and \rfrs{dodtPSeq}{detdtPSeq} describe genuine secular trends.
\subsection{Polar orbit}\lb{pola}
Let us assume that the body's spin axis, irrespectively of its orientation in the adopted coordinate system, i.e., for generic values of $\alpha,\,\delta$, lies somewhere in the satellite's orbital plane between $\bds{\hat{l}}$ and $\bds{\hat{m}}$. Such an orbital geometry is widely adopted in spacecraft-based missions to solar system planets like, e.g., Juno at Jupiter \citep{2017SSRv..213....5B}. In such a scenario, it is
\begin{align}
\kl \lb{pel}& \neq 0, \\ \nonumber \\
\km \lb{pm}& \neq 0, \\ \nonumber \\
\kh \lb{ph}& =0.
\end{align}
\Rfr{ph} implies that the orbital angular momentum is perpendicular to the body's spin axis.
\Rfrs{kl}{kh} tell us that the conditions of \rfrs{pel}{ph} are satisfied for
\begin{align}
I \lb{inc}& = \rp{\pi}{2}, \\ \nonumber \\
\Omega \lb{omg}& = \alpha;
\end{align}
indeed, with \rfrs{inc}{omg}, one has just
\begin{align}
\kl & =\cos\delta, \\ \nonumber \\
\km & =\sin\delta, \\ \nonumber \\
\kh & =0.
\end{align}

As a consequence, \rfrs{dadtBS}{detdtPS} reduce to
\begin{align}
\dert a t \lb{dadtBSpl}& = \rp{9~e^2~\ton{6 + e^2}~\nk~J_2~\mu~R_\mathrm{e}^2~\sin\qua{2~\ton{\delta - \omega}}}{8~c^2~a^2~\ton{1-e^2}^4}, \\ \nonumber \\
\dert e t \lb{dedtBSpl}& = \rp{21~e~\ton{2 + e^2}~\nk~J_2~\mu~R_\mathrm{e}^2~\sin\qua{2~\ton{\delta - \omega}}}{16~c^2~a^3~\ton{1-e^2}^3}, \\ \nonumber\\
\dert I t \lb{dIdtBSpl}& = 0,\\ \nonumber \\
\dert \Omega t \lb{dOdtBSpl}& = 0,\\ \nonumber \\
\dert \omega t \lb{dodtBSpl}& = -\rp{3~\nk~J_2~\mu~R_\mathrm{e}^2~\grf{-8 + 3~e^2 + 14~\cos\qua{2~\ton{\delta - \omega}}} }{
16~c^2~a^3~\ton{1 - e^2}^3},\\ \nonumber \\
\dert \eta t \lb{detdtBSpl}& = \rp{\nk~J_2~\mu~R_\mathrm{e}^2~\grf{80 + 73~e^2 + 42~\ton{1 + 2~e^2}~\cos\qua{2~\ton{\delta - \omega}}} }{
16~c^2~a^3~\ton{1 - e^2}^{5/2}},
\end{align}
while \rfrs{dadtPS}{detdtPS} become
\begin{align}
\dert a t \lb{dadtPSpl}& = 0, \\ \nonumber \\
\dert e t \lb{dedtPSpl}& = 0, \\ \nonumber\\
\dert I t \lb{dIdtPSpl}& = -\rp{9~G~S~R^2_\mathrm{e}~\varepsilon^2~\cos\delta~\grf{4 + 6~e^2 + 5~e^2~\cos\qua{2\ton{\delta - \omega}}}}{56~c^2~a^5~\ton{1-e^2}^{7/2}},\\ \nonumber \\
\dert \Omega t \lb{dOdtPSpl}& = -\rp{9~G~S~R^2_\mathrm{e}~\varepsilon^2~\sin\delta~\grf{4 + 6~e^2 + 5~e^2~\cos\qua{2\ton{\delta - \omega}}}}{56~c^2~a^5~\ton{1-e^2}^{7/2}},\\ \nonumber \\
\dert \omega t \lb{dodtPSpl}& = 0,\\ \nonumber \\
\dert \eta t \lb{detdtPSpl}& = 0.
\end{align}
\Rfrs{dadtBSpl}{dedtBSpl}, \rfrs{dodtBSpl}{detdtBSpl}, and  \rfrs{dIdtPSpl}{dOdtPSpl}, in addition to secular trends, when present, include also  long-period signatures due to the evolution of pericentre which is mainly driven by the zonal harmonics of the Newtonian component of the multipolar field of the central body \citep{Capde05}.

It can be shown that, \textcolor{black}{for the orbital configuration of} \rfrs{inc}{omg}, the classical precessions of $I$ and $\Omega$ due to the zonal harmonics of the primary's gravitational potential vanish\textcolor{black}{; see Appendix A of \citet{2023MNRAS.523.3595I}}. It is important  since such classical effects are usually regarded as a major source of systematic bias in assessing the error budget of a mission dedicated to measure certain general relativistic features of motion.

\textcolor{black}{
On the other hand, \rfr{dadtBS} has not a classical counterpart acting as a systematic bias: it is true for any orbital configuration and primary's spin axis orientation, being the Newtonian oblateness-driven rate of change of $a$ exactly zero. As far as the eccentricity is concerned, in general, the odd zonal harmonics of the classical gravity field of the source induce net variations per orbit which do not vanish for the polar geometry of \rfrs{inc}{omg}.
It turns out that they are harmonic signatures oscillating with odd multiples of the pericentre frequency, while \rfr{dedtBSpl} varies just at twice the frequency of $\omega$. More specifically, by following the approach outlined in Appendix A of \citet{2023MNRAS.523.3595I}, one has
\begin{align}
\dot e^{\ton{J_3}}\lb{kazzo3} &= \rp{3\,\nk\,R_\mathrm{e}^3\,J_3\,\sin\ton{\delta - \omega}}{8\,a^3\,\ton{1 - e^2}^2}, \acap
\dot e^{\ton{J_5}} &= \rp{15\,\nk\,R_\mathrm{e}^5\,J_5\,\grf{8 + 13\,e^2 + 14\,e^2\,\cos\qua{2 \ton{\delta - \omega}}}\,\sin\ton{\delta - \omega}}{256\,a^5\,\ton{1 - e^2}^4},\acap
\dot e^{\ton{J_7}} &= \rp{105\,\nk\,R_\mathrm{e}^7\,J_7\,\grf{
40 + 208\,e^2 + 82\,e^4 + 6\,e^2\,\ton{36 + 19\,e^2}\,\cos\qua{2\ton{\delta - \omega}} + 33\,e^4\,\cos\qua{4\ton{\delta - \omega}}
}\,\sin\ton{\delta - \omega}}{8192\,a^7\,\ton{1 - e^2}^6},
\end{align}
and so on.
Either the even and the odd zonals make the pericentre to vary, as per Equations (A10) to (A16) of \citet{2023MNRAS.523.3595I}.
}
Furthermore, if the condition
\eqi
\omega = \delta -\rp{\pi}{2}\lb{blah}
\eqf
is selected, it turns out that the pericentre precessions due to the odd zonals vanish \citep{2023MNRAS.523.3595I}. On the other hand, if the condition of \rfr{blah} holds, \rfrs{dadtBSpl}{dedtBSpl} vanish as well.
%
%
\section{The Jovian scenario for a dedicated mission}\lb{jup}
In view of its size, Jupiter, whose relevant physical parameters are listed in Table\,\ref{tab1}, seems to be the ideal\footnote{The fifth planet of our solar system was often been considered for testing various aspects of gravitomagnetism over the years; see, e.g., \citet{1975Ap&SS..32....3H,2000CQGra..17.2399M,2000CQGra..17.2381T,2000CQGra..17..783T,2010NewA...15..554I,2017FrASS...4...11S,2019MNRAS.484.4811I}.} candidate, at least in principle, to try to measure the general relativistic effects treated in the previous sections with a dedicated spacecraft-based mission. With a pinch of irony, we preliminarily dub it IORIO, acronym of In-Orbit Relativity Iuppiter\footnote{\textit{Iupp\u{\i}t\u{e}r} is one of the forms of the Latin noun of the god Jupiter.} Observatory, or, equally well, of
IOvis\footnote{In Latin, \textit{I\u{o}vis} means ``of Jupiter".} Relativity In-orbit Observatory \citep{2019MNRAS.484.4811I,2019MNRAS.485.4090I}.
\begin{table}[ht!]
\caption{Relevant physical parameters of Jupiter \citep{2003AJ....126.2687S,2007CeMDA..98..155S,iers10,2018Natur.555..220I}.
}\lb{tab1}
\begin{center}
\begin{tabular}{|l|l|l|}
  \hline
Parameter  & Units & Numerical Value \\
\hline
$\mu$ & $\textrm{m}^3~\textrm{s}^{-2}$ & $1.26713\times 10^{17}$ \citep{iers10}\\
$J_2$ & $\times 10^{-6}$ & $14696.572$ \citep{2018Natur.555..220I}\\
\textcolor{black}{$J_3$} & \textcolor{black}{$\times 10^{-6}$} & \textcolor{black}{$-0.042$} \citep{2018Natur.555..220I}\\
$S$ & $\textrm{kg~m}^2$~$\textrm{s}^{-1}$ & $6.9\times 10^{38}$ \citep{2003AJ....126.2687S}\\
$\alpha$ & $\textrm{deg}$ & $268.057132$ \citep{2018Natur.555..220I}\\
$\delta$ & $\textrm{deg}$ & $64.497159$ \citep{2018Natur.555..220I}\\
$R_\mathrm{e}$ & $\textrm{km}$ & $71492$ \citep{2007CeMDA..98..155S}\\
$R_\mathrm{p}$ & $\textrm{km}$ & $66854$ \citep{2007CeMDA..98..155S}\\
\hline
$\varepsilon$ & $-$ & $0.354$\\
\hline
\end{tabular}
\end{center}
\end{table}
Moreover, both for typical planetological goals\footnote{Polar orbits, taking the spacecraft over the poles of the planet, allow the probe to traverse the latter in a north--south direction; they are optimal for mapping and monitoring it \citep{2000Monte}.} and to preserve the onboard scientific instrumentation from harmful radiation\footnote{Radiation belts are the regions of a magnetosphere where high energy charged particles, such as electrons, protons, and heavier ions, are trapped in large amounts. All planets in our solar system having a sufficiently intense magnetic field, such as Earth, Jupiter, Saturn, Uranus, and Neptune, host radiation belts \citep{2010JGRA..11512220M,2014JGRA..119.9729M}. Jupiter has the most complex and energetic radiation belts in our solar system and one of the most challenging space environments to measure and characterize in-depth \citep{2022ExA....54..745R}.}, the orbits of the spacecraft targeted to the fifth planet of our solar system are usually just polar and widely elliptical; thus, we can rely upon the results of Section\,\ref{pola}. A notable existing example of such a peculiar orbital geometry is provided by the ongoing mission Juno \citep{2017SSRv..213....5B}. It is currently orbiting Jupiter along a wide 53 days orbit characterized by a pericenter height of
\eqi
h_\mathrm{peri}=4200\,\mathrm{km}\lb{hperi}
\eqf
and an apocenter height of
\eqi
h_\mathrm{apo}=8.1\times 10^6\,\mathrm{km},\lb{hapo2}
\eqf
corresponding to a semimajor axis
\eqi
a =4.1\times 10^6 \,\mathrm{km} = 57.7\,R_\mathrm{e}\lb{sma}
\eqf
and an eccentricity
\eqi
e = 0.981.\lb{ecc}
\eqf
Originally, a 13 days orbit, corresponding to a lower apocenter height of
\eqi
h_\mathrm{apo} = 3.2\times 10^6\,\mathrm{km},\lb{hapo1}
\eqf
and eccentricity
\eqi
e = 0.954
\eqf
was planned for Juno \citep{2007AcAau..61..932M}, but problems with two helium valves in its propulsion system in 2016 October prevented to meet such a goal\footnote{See \url{https://www.science.org/content/article/avoid-risk-misfire-nasas-juno-probe-will-keep-its-distance-jupiter}.}.

For the sake of concreteness, in the following we will consider a putative spacecraft having the same pericenter height as Juno, given by \rfr{hperi}, while its apocenter height is allowed to vary from, say,
\eqi
h_\mathrm{apo} = 1.5\times 10^6\,\mathrm{km}\lb{hapo3}
\eqf
to the current Juno value of \rfr{hapo2}, corresponding to a range variation for $e$ of $\simeq 0.92-0.98$.
Obtaining \rfr{hapo3} would be a demanding task since more fuel would be needed, and the spacecraft would be exposed to a larger amount of radiation.

\begin{figure}[ht!]
\centering
\centerline{
\vbox{
\begin{tabular}{c}
\epsfxsize= 12.5 cm\epsfbox{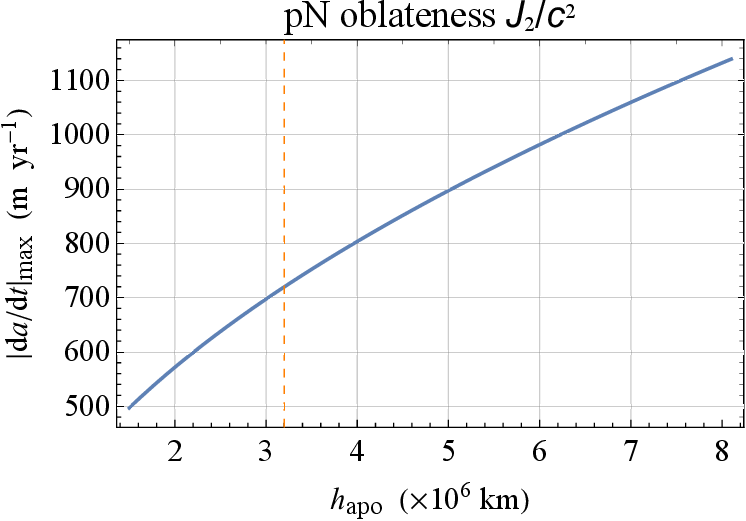} \\ \epsfxsize= 12.5 cm\epsfbox{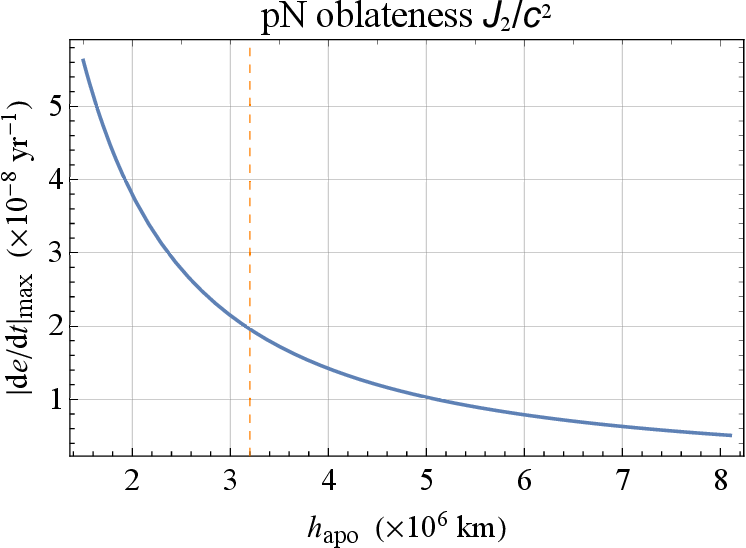}\\
\end{tabular}
}
}
\caption{
Plot of the amplitudes, in $\mathrm{m}\,\mathrm{yr}^{-1}$ and $10^{-8}\,\mathrm{yr}^{-1}$, respectively, of \rfrs{dadtBSpl}{dedtBSpl}, calculated for the polar orbital configuration of \rfrs{inc}{omg}, as functions of the apocenter height $h_\mathrm{apo}$ for a fixed value of the pericenter height $h_\mathrm{peri} = 4200\, \mathrm{km}$. The dashed vertical line corresponds to $h_\mathrm{apo} = 3.2\times 10^6\,\mathrm{km}$ which was the originally planned apocenter height of Juno before the problems encountered by its propulsion system.
}\label{fig1}
\end{figure}

\begin{figure}[ht!]
\centering
\centerline{
\vbox{
\begin{tabular}{c}
\epsfxsize= 12.5 cm\epsfbox{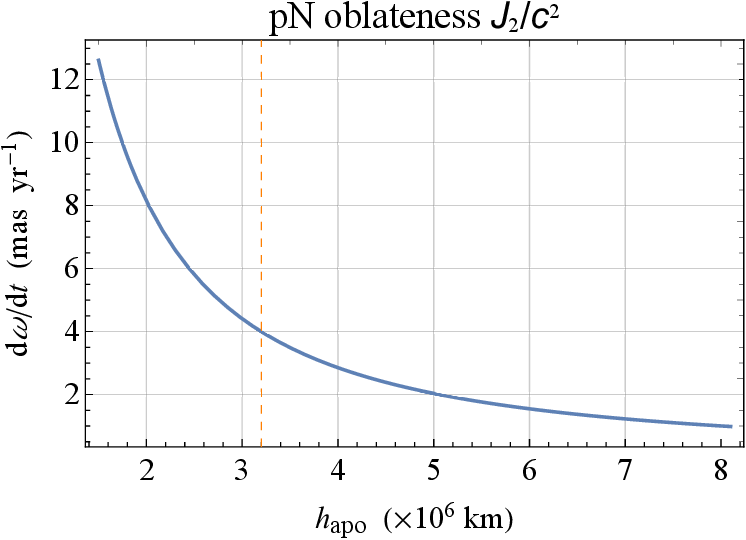} \\ \epsfxsize= 12.5 cm\epsfbox{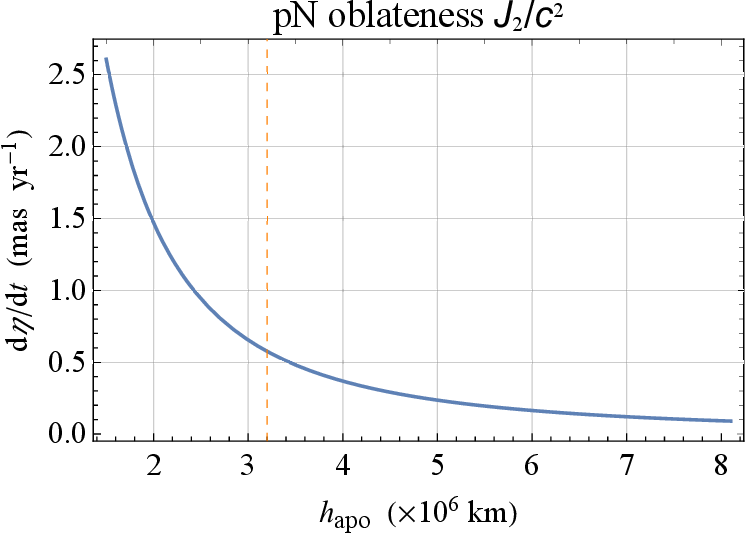}\\
\end{tabular}
}
}
\caption{
Plot  of \rfrs{dodtBSpl}{detdtBSpl}, in $\mathrm{mas}\,\mathrm{yr}^{-1}$, calculated for the polar orbital configuration of \rfrs{inc}{omg} with the condition of \rfr{blah}, as functions of the apocenter height $h_\mathrm{apo}$ for a fixed value of the pericenter height $h_\mathrm{peri} = 4200\, \mathrm{km}$. The dashed vertical line corresponds to $h_\mathrm{apo} = 3.2\times 10^6\,\mathrm{km}$ which was the originally planned apocenter height of Juno before the problems encountered by its propulsion system.
}\label{fig2}
\end{figure}

\begin{figure}[ht!]
\centering
\centerline{
\vbox{
\begin{tabular}{c}
\epsfxsize= 12.5 cm\epsfbox{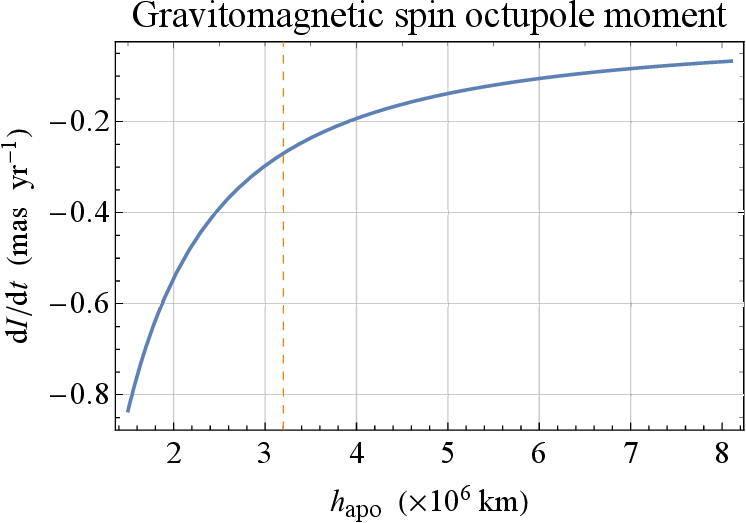} \\ \epsfxsize= 12.5 cm\epsfbox{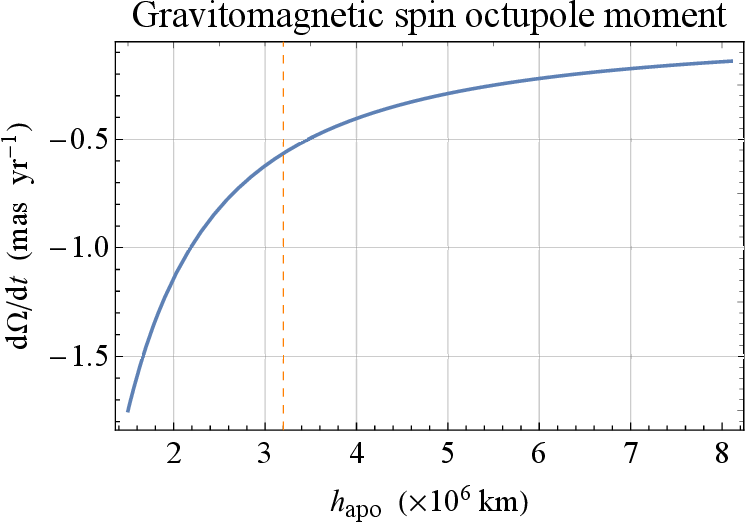}\\
\end{tabular}
}
}
\caption{
Plot  of \rfrs{dIdtPSpl}{dOdtPSpl}, in $\mathrm{mas}\,\mathrm{yr}^{-1}$, calculated for the polar orbital configuration of \rfrs{inc}{omg} with the condition of \rfr{blah}, as functions of the apocenter height $h_\mathrm{apo}$ for a fixed value of the pericenter height $h_\mathrm{peri} = 4200\, \mathrm{km}$. The dashed vertical line corresponds to $h_\mathrm{apo} = 3.2\times 10^6\,\mathrm{km}$ which was the originally planned apocenter height of Juno before the problems encountered by its propulsion system.
}\label{fig3}
\end{figure}

Figures\,\ref{fig1} to \ref{fig2} deal with the pN gravitoelectric effects of \rfrs{dadtBSpl}{detdtBSpl}, while Figure\,\ref{fig3} depicts the gravitomagnetic precessions of \rfrs{dadtPSpl}{detdtPSpl}. The conditions of \rfrs{inc}{omg} and of \rfr{blah} were adopted in the calculation in order to obtain Figures\,\ref{fig2} to \ref{fig3}.

According to Figure\,\ref{fig1}, the amplitude of the $J_2/c^2$ long-period signature of $a$ ranges from $500\,\mathrm{m\,yr}^{-1}=0.01\,\mathrm{mm\,s}^{-1}$ to $1100\,\mathrm{m\,yr}^{-1}=0.03\,\mathrm{mm\,s}^{-1}$, a seemingly unexpected feature due to the impact of $e$ in \rfr{dadtBSpl} for increasingly larger values of it. Such figures are not far from the two-way \textit{Ka}-band range rate residuals $\Delta\dot\rho$ of Juno displayed in \citet{2018Natur.555..220I} whose ranges of variation amount to $0.050\,\mathrm{mm\, s}^{-1}$, with a root-mean-square value of
$0.015\,\mathrm{mm\, s}^{-1}$. However, caution is in order when such risky comparisons are made since the directly measurable range-rate and the theoretically computed semimajor axis are quite distinct quantities.
\textcolor{black}{As far as the $J_2/c^2$ effect on $e$ is concerned, it turns out that, according to Table\,\ref{tab1}, the nominal value of \rfr{kazzo3}, which is a potentially major source of systematic bias, is smaller than it by about one order of magnitude. }
The $J_2/c^2$ pericentre precession, shown in Figure\,\ref{fig2}, is in the range $\simeq 2-12$ milliarcseconds per year ($\mathrm{mas\,yr}^{-1}$). It would be overwhelmed by the classical secular trend due to $J_2$. Furthermore, both the latter and the 1pN signature are proportional to $J_2$ itself; their ratio is independent of it, so that one may not invoke any future improvement in our knowledge of the first even zonal of the Jovian gravity field   to reduce the impact of the Newtonian effect on the 1pN one. \textcolor{black}{A further potentially major source of systematic bias is represented by the competing  classical $N$-body perturbations due to the Galilean moons of Jupiter, which orbit in its equatorial plane. It preliminarily turns out that\footnote{\textcolor{black}{See Equations (1)-(5) of \citet{2020ApJ...889..152I} in which the parameters of the perturber are marked with $\bullet$.}}, if on the one hand, they leave the semimajor axis unaffected, on the other hand, for a polar orbit, they induce nonvanishing doubly-averaged orbital disturbances on $e$ and $\omega$. According to the present-day level of uncertainty\footnote{\textcolor{black}{See \url{https://ssd.jpl.nasa.gov/sats/phys\textunderscore par/} on the Internet.}} in their masses as per the Planetary Satellite Ephemeris JUP365 \citep{jup365}, the resulting mismodelled signatures of $e$ and $\omega$ would be roughly one order of magnitude larger than, or about of the same order of magnitude of, the $J_2/c^2$ ones, apart for Callisto for which they are smaller than the pN ones.
In the near future, the masses of the three outer Galilean
satellites will be accurately determined by the JUpiter ICy moons Explorer  \citep[JUICE; ]{2013P&SS...78....1G} and Clipper missions \citep{2022EGUGA..24.6052K}, while the flybys of Io by Juno should allow to improve the mass of Io as well. In particular, according to Tables 1–3
of \citet{2021AeMiS.100..195M}, the masses of Europa, Ganymede, and Callisto should be determined by JUICE with an improvement of about 1–2 orders of magnitude with respect to the errors by \citet{jup365}.
}

Figure\,\ref{fig3} tells us that the nonvanishing gravitomagnetic spin octupole orbital effects  amount to $\simeq 0.1-2\,\mathrm{mas\,yr}^{-1}$. \textcolor{black}{The impact of the Galilean moons of Jupiter would be of no concern for \rfr{dOdtPSpl} if viewed in a frame\footnote{\textcolor{black}{In it, \rfr{dIdtPSpl} vanishes since $\delta=\pi/2$.}} aligned with the Jovian equator since the $N-$body node rate of change of a polar orbit acted upon by an equatorial perturber vanishes, as per Equation (4) of \citet{2020ApJ...889..152I}. If, instead, the ICRF is adopted, the nonvanishing inclination of the orbits of the Galilean satellites to the Celestial Equator would make that the amplitudes of their mismodelled perturbations on $I$ and $\Omega$ \citep[Equations (4)-(5)]{2020ApJ...889..152I} may be up to about one order of magnitude larger than \rfrs{dIdtPSpl}{dOdtPSpl}. The Sun does not represent an issue since it turns out that, by assuming $\sigma_{\mu_\odot}= 1\times 10^{10}\,\mathrm{m^3\,s^{-2}}$ \citep{iers10}, its mismodelled classical perturbations are several orders of magnitude smaller than the pN effects of interest.}

\begin{figure}[ht!]
\centering
\centerline{
\vbox{
\begin{tabular}{c}
\epsfxsize= 12.5 cm\epsfbox{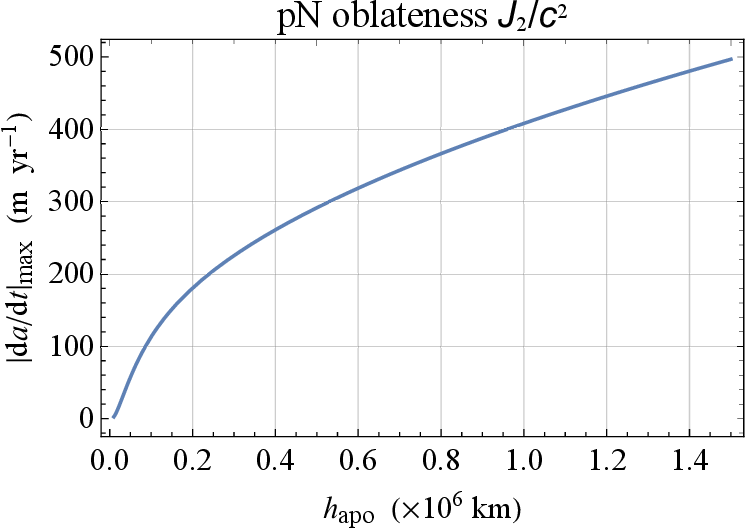} \\ \epsfxsize= 12.5 cm\epsfbox{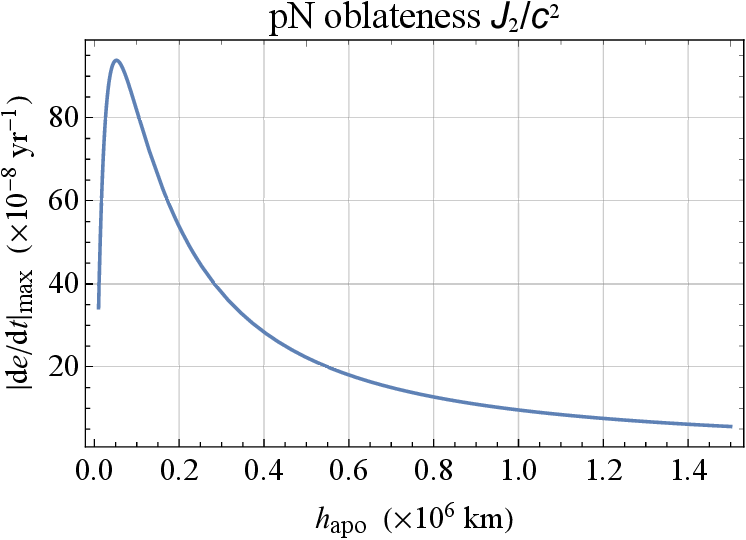}\\
\end{tabular}
}
}
\caption{
Plot of the amplitudes, in $\mathrm{m}\,\mathrm{yr}^{-1}$ and $10^{-8}\,\mathrm{yr}^{-1}$, respectively, of \rfrs{dadtBSpl}{dedtBSpl}, calculated for the polar orbital configuration of \rfrs{inc}{omg}, as functions of the apocenter height $h_\mathrm{apo}$ for a fixed value of the pericenter height $h_\mathrm{peri} = 4200\, \mathrm{km}$.
}\label{fig4}
\end{figure}

\begin{figure}[ht!]
\centering
\centerline{
\vbox{
\begin{tabular}{c}
\epsfxsize= 12.5 cm\epsfbox{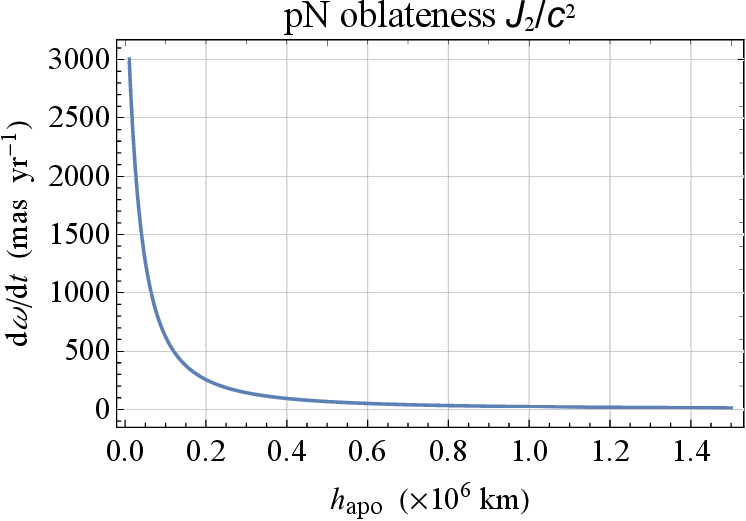} \\ \epsfxsize= 12.5 cm\epsfbox{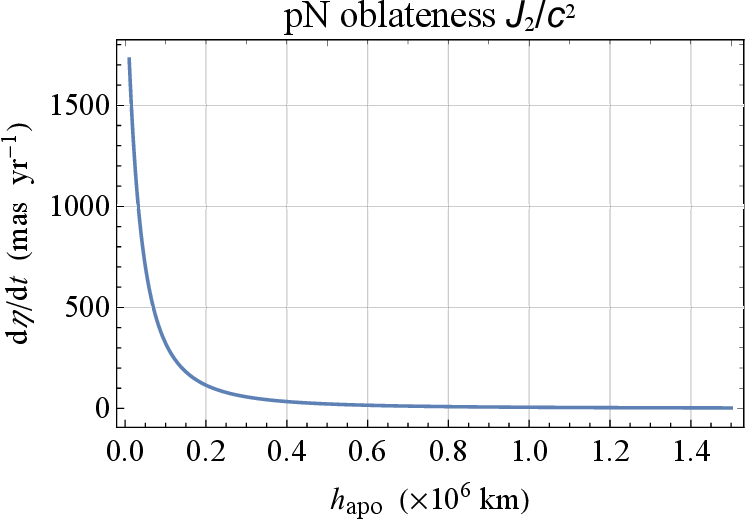}\\
\end{tabular}
}
}
\caption{
Plot  of \rfrs{dodtBSpl}{detdtBSpl}, in $\mathrm{mas}\,\mathrm{yr}^{-1}$, calculated for the polar orbital configuration of \rfrs{inc}{omg} with the condition of \rfr{blah}, as functions of the apocenter height $h_\mathrm{apo}$ for a fixed value of the pericenter height $h_\mathrm{peri} = 4200\, \mathrm{km}$.
}\label{fig5}
\end{figure}

\begin{figure}[ht!]
\centering
\centerline{
\vbox{
\begin{tabular}{c}
\epsfxsize= 12.5 cm\epsfbox{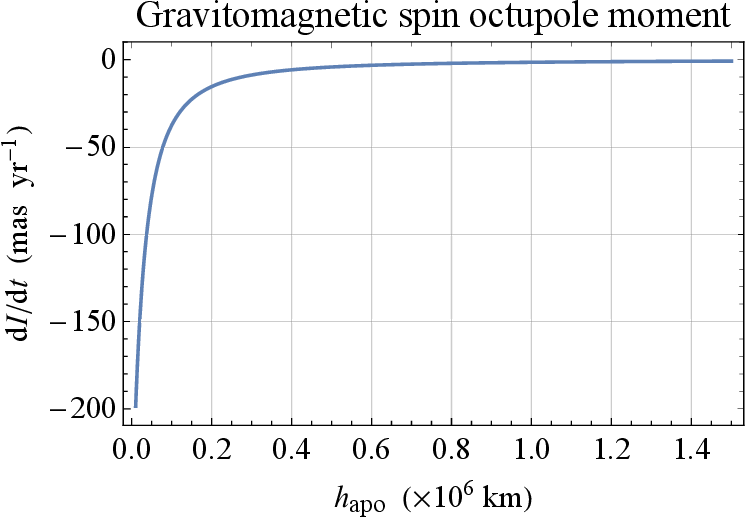} \\ \epsfxsize= 12.5 cm\epsfbox{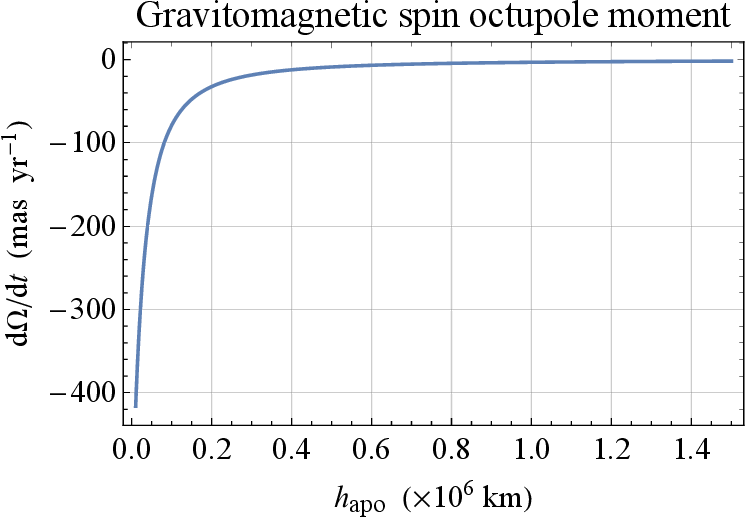}\\
\end{tabular}
}
}
\caption{
Plot  of \rfrs{dIdtPSpl}{dOdtPSpl}, in $\mathrm{mas}\,\mathrm{yr}^{-1}$, calculated for the polar orbital configuration of \rfrs{inc}{omg} with the condition of \rfr{blah}, as functions of the apocenter height $h_\mathrm{apo}$ for a fixed value of the pericenter height $h_\mathrm{peri} = 4200\, \mathrm{km}$.
}\label{fig6}
\end{figure}

Despite inserting a spacecraft into a moderately elliptical orbits around Jupiter is a very daunting task because of the exceedingly amount of fuel required, we deem the study of the relativistic effects even in such an unlikely scenario worthy of investigation. By exploring the eccentricity range $e\simeq 0.05-0.9$, with the pericenter height fixed to the value of \rfr{hperi}, yields the remarkable relativistic signatures of Figures\,\ref{fig4} to \ref{fig6}.
\clearpage
In this case, the $J_2/c^2$ precessions can reach even the arcsec yr$^{-1}$ level, while the gravitomagnetic signatures can be as large as a few hundred mas yr$^{-1}$. As far as the semimajor axis is concerned, the amplitude of its $J_2/c^2$ signature
ranges from about $3\,\mathrm{m\,yr}^{-1}=3\times 10^{-4}\,\mathrm{mm\,s}^{-1}$ to $500\,\mathrm{m\,yr}^{-1}=0.05\,\mathrm{mm\,s}^{-1}$. \textcolor{black}{Now, the mismodelled $N-$ body orbital perturbations due to all the Galilean moons turn out to  be negligible.}

Such results  demonstrate that, perhaps, it would be worth trying to actually carry out such a mission, however challenging it is. After all, it should not be neglected that, in addition to the relatively tiny pN effects induced by the multipoles of Jupiter, it could measure also the traditional and much larger pN mass monopole and spin-dipole effects.

\textcolor{black}{Incidentally, for, say, a Jovicentric Juno-like orbit, the aforementioned mixed effects of order $\mathcal{O}\ton{J_2^2/c^2}$ and $\mathcal{O}\ton{S\,\varepsilon^2\,J_2/c^2}$ would be smaller than the direct ones retrievable in Figures\,\ref{fig1} to \ref{fig3} by a factor $\simeq J_2\,\ton{R_\mathrm{e}/a}^2 = 4\times 10^{-6}$, as per \rfr{sma} and Table\,\ref{tab1}.}

About the possibility of looking at Saturn as well, its relevant physical parameters are listed in Table\,\ref{tab2}.
\begin{table}[ht!]
\caption{Relevant physical parameters of Saturn \citep{2003AJ....126.2687S,2007CeMDA..98..155S,2022AJ....164..199J}.
The value for the ellipticity $\varepsilon$ is obtained from the figures reported for $R_\mathrm{e}$ and $R_\mathrm{p}$.}\lb{tab2}
\begin{center}
\begin{tabular}{|l|l|l|}
  \hline
Parameter  & Units & Numerical value \\
\hline
$\mu$ & $\textrm{m}^3~\textrm{s}^{-2}$ & $3.79312\times 10^{16}$ \citep{2022AJ....164..199J}\\
$J_2$ & $\times 10^{-6}$ & $16290.615$ \citep{2022AJ....164..199J}\\
$S$ & $\textrm{kg~m}^2$~$\textrm{s}^{-1}$ & $1.4\times 10^{38}$ \citep{2003AJ....126.2687S}\\
$\alpha$ & $\textrm{deg}$ & $40.594872$ \citep{2022AJ....164..199J}\\
$\delta$ & $\textrm{deg}$ & $83.534351$ \citep{2022AJ....164..199J}\\
$R_\mathrm{e}$ & $\textrm{km}$ & $60268$ \citep{2007CeMDA..98..155S}\\
$R_\mathrm{p}$ & $\textrm{km}$ & $54364$ \citep{2007CeMDA..98..155S}\\
\hline
$\varepsilon$ & $-$ & $0.431$\\
\hline
\end{tabular}
\end{center}
\end{table}
From them and from Table\,\ref{tab1}, it turns out that, despite the Kronian oblateness and ellipticity are slightly larger than the Jovian ones,  the relativistic effects considered so far, with the same orbit, are larger for Jupiter by about one order of magnitude. Indeed, it is
\begin{align}
\rp{M^\mathrm{Sat}\,J_2^\mathrm{Sat}\,\ton{R_\mathrm{e}^\mathrm{Sat}}^2}{M^\mathrm{Jup}\,J_2^\mathrm{Jup}\,\ton{R_\mathrm{e}^\mathrm{Jup}}^2} & \simeq 0.23, \acap
\rp{S_\mathrm{Sat}\,\ton{R_\mathrm{e}^\mathrm{Sat}}^2\,\varepsilon^2_\mathrm{Sat}}{S_\mathrm{Jup}\,\ton{R_\mathrm{e}^\mathrm{Jup}}^2\,\varepsilon^2_\mathrm{Jup}}&\simeq 0.21.
\end{align}
5
\section{Probing the extended mass component in Sgr A$^\ast$}\lb{SgrA}
The general validity of \rfrs{dadtBS}{detdtBS} and of \rfrs{dadtPS}{detdtPS} allows one to apply them, in principle,
also to the highly elliptical paths of the S stars \citep{2020ApJ...896..100A} orbiting the supermassive black hole (SMBH) in Sgr A$^\ast$ at the Galactic Center \citep{2008ApJ...689.1044G,2010RvMP...82.3121G} in order, e.g., to characterize the predicted mass component enclosed by the stellar orbits\footnote{Until now, only the S2 star has been used to this aim; see, e.g., \citet{2001A&A...374...95R,2007PASP..119..349N,2007PhRvD..76f2001Z,2015PhyU...58..772D,2015JETPL.101..777D,2022A&A...660A..13H,2022MNRAS.511L..35A,2022A&A...657L..12G}.} \citep{1999PhRvL..83.1719G,2005AN....326...83M,2009ApJ...692.1075G,2022NatSR..1215258C}. Such an extended dark mass distribution surrounding Sgr A$^\ast$ might be made of faint S stars, neutron stars, stellar mass black holes, faint accretion gas clouds, stellar remnants and nonbaryonic dark matter as well. Should it has  departures from spherically symmetry, they may be dynamically probed, in principle, with the pN orbital effects investigated here.

As far as the SMBH itself is concerned, its direct effects are too small to be of any relevance. As an example, with the replacement
\eqi
J_2 \rightarrow -\rp{Q_2}{M~R_\mathrm{e}^2}
\eqf
in \rfr{dadtBS},
by using\footnote{It is a manifestation of the celebrated \virg{no-hair theorem} \citep{1971PhRvL..26..331C,1975PhRvL..34..905R} according to which
the mass $\mathcal{M}_\bullet^\ell$ and the spin $\mathcal{S}_\bullet^\ell$ moments of a Kerr black hole \citep{1970JMP....11.2580G,1974JMP....15...46H} are connected by the relation $\mathcal{M}_\bullet^\ell + i\,\mathcal{S}_\bullet^\ell= M_\bullet \ton{i\,S_\bullet/c\,M_\bullet}^\ell$, where $i=\sqrt{-1}$. The odd mass moments and even spin moments
are identically zero; $\mathcal{M}_\bullet^2$ is the quadrupole mass moment, corresponding just to \rfr{Q2}.
} \citep{1971PhRvL..26..331C,1975PhRvL..34..905R}
\eqi
Q^\bullet_2 = -\rp{S_\bullet^2}{c^2~M_\bullet}\lb{Q2}
\eqf
and \citep{1963PhRvL..11..237K,2015CQGra..32l4006T}
\eqi
S_\bullet = \chi_g\rp{M_\bullet^2~G}{c},
\eqf
the amplitude of \rfr{dadtBS} can be cast into the form
\eqi
\dert a t \propto -\rp{9~e^2~\ton{6 + e^2}~\pi^3~\chi^2_g~\mu_\bullet^2}{c^6~\ton{1 - e^2}^4~\Pb^3}.\lb{bh}
\eqf
In the case of the recently discovered star S4716, characterized by $\Pb=4.02\,\mathrm{yr},\,e=0.756$ \citep{2022ApJ...933...49P}, \rfr{bh} yields\footnote{The values $\chi_g = 0.5,\,M_\bullet = 4.1\times 10^6\,M_\odot$ \citep{2020ApJ...899...50P,2022ApJ...933...49P} were
used in \rfr{bh}.}
\eqi
\left|\dert a t\right|\propto 7\times 10^{-11}\,\mathrm{m\,yr}^{-1}.
\eqf

\section{Summary and conclusions}\lb{fine}
We analytically worked out the pN orbital effects induced on a test particle by the quadrupole mass moment and the gravitomagnetic spin octupole moment of an axisymmetric rotating body. The resulting explicit expressions of the long-term rates of change of the satellite's osculating Keplerian orbital elements, averaged over one orbital period, retain a general validity since they hold for any orbital configuration and for an arbitrary orientation of the body's spin axis in space. In general, they are all nonzero, with the exception of the gravitomagnetic variation of the semimajor axis which vanishes over an orbital revolution.

Our results were subsequently specialized to two particular orbital geometries: (a) an equatorial orbit; (b) a polar orbit.
In the scenario a), only the pericenter and the mean anomaly at epoch undergo nonzero net effects for both the pN accelerations considered; such orbital features of motion turn out to be genuine secular trends. The scenario b) is more complex. Indeed, in the case of the pN oblateness, only the inclination and the node stay constant. While the semimajor axis and the eccentricity experience purely long-period variations because of the generally varying pericenter entering the expressions of their averaged rates of change, the pericenter and the mean anomaly at epoch are affected also by secular trends in addition to long-period signals. It should be recalled that, in a realistic scenario, the pericenter does change mainly because of the zonal harmonics of the Newtonian part of the multipolar expansion of the gravitational potential of the central body. As far as the gravitomagnetic octupolar field is concerned, the only orbital elements that vary, on average, are the inclination and the node; they experience both secular and long-period effects.

The case b) was applied to a hypothetical scenario around Jupiter by first adopting the same pericenter height of the current Juno spacecraft and varying the apocenter height in such a way that the eccentricity ranged from $e = 0.92$ to the Juno's present value  $e = 0.98$. The resulting amplitude of the long-period pN signature of the semimajor axis due to the Jovian oblateness is comprised within $500\,\mathrm{m\,yr}^{-1}=0.01\,\mathrm{mm\,s}^{-1}\,\ton{e=0.92}$ and $1100\,\mathrm{m\,yr}^{-1}=0.03\,\mathrm{mm\,s}^{-1}\,\ton{e=0.98}$. Although the following improper comparison is potentially misleading, it may be interesting to cautiously noting that the present-day accuracy level in measuring the range-rate shift of Juno is just at the $\simeq 0.01\,\mathrm{mm\,s}^{-1}$ level. The other nonvanishing pN oblateness-driven effects on the pericenter and the mean anomaly at epoch amount to about $2-12\,\mathrm{mas\,yr}^{-1}$, while the gravitomagnetic rates of the inclination and the node are at the  $\simeq 0.1-2\,\mathrm{mas\,yr}^{-1}$ level. Then, much less elliptical orbits with the same pericenter of Juno were considered, although they are at present almost impossible to achieve practically. In this case,  the pN gravitoelectric rates can be as large as $\simeq 1-3\,\mathrm{arcsec\, yr}^{-1}$, while the spin octupole effects can reach the level of a few hundred mas yr$^{-1}$. Given the same orbital configuration, it turns out that the aforementioned effects around Saturn would be about an order of magnitude smaller.

Another potentially viable scenario for our results is represented by the highly elliptical orbits of the S stars moving around the supermassive black hole in Sgr A$^\ast$ at the Galactic Center. Indeed, it is likely surrounded by an extended matter distribution which can be made either of nonbaryonic dark matter or by the remnants of tidally disrupted stars, pulsars, etc. Such a halo is not necessarily spherically symmetric, and, in principle, the orbital effects calculated here may be useful to get more information on it.
%
\begin{appendices}
\section{Notations and definitions}\lb{appenA}
Here, some basic notations and definitions used throughout the text are presented \citep{Sof89,1991ercm.book.....B,2003ASSL..293.....B,2011rcms.book.....K,2014grav.book.....P,SoffelHan19}.
\begin{itemize}
\item[] $G:$ Newtonian constant of gravitation
\item[] $c:$ speed of light in vacuum
\item[] $M:$ mass of the central body
\item[] $M_\odot:$ mass of the Sun
\item[] $\mu:=  GM:$ gravitational parameter of the central body
\item[] $\bds S:$ angular momentum of the central body
\item[] $S:$ magnitude of the angular momentum of central body
\item[] $\chi_g:$ dimensionless spin parameter of a Kerr black hole: it is $\left|\chi_g\right|\leq 1$.
\item[] ${\bds{\hat{k}}} = \grf{{\hat{k}}_x,~{\hat{k}}_y,~{\hat{k}}_z}:$ spin axis of the central body with respect to some reference frame
\item[] $\alpha:$ longitude of the spin axis of the central body in some reference frame
\item[] $\delta:$ latitude of the spin axis of the central body in some reference frame
\item[] ${\hat{k}}_x = \cos\alpha\,\cos\delta:$ $x$ component of the  spin axis of the central body with respect to some reference frame
\item[] ${\hat{k}}_y = \sin\alpha\,\cos\delta:$ $y$ component of the  spin axis of the central body with respect to some reference frame
\item[] ${\hat{k}}_z = \sin\delta:$ $z$ component of the  spin axis of the central body with respect to some reference frame
\item[] $R_\textrm{e}:$ equatorial radius of the central body
\item[] $R_\textrm{p}:$ polar radius of the central body
\item[] $\varepsilon:= \sqrt{1 - \ton{\rp{R_\textrm{p}}{R_\textrm{e}}}^2}:$ ellipticity of the central body
\item[] $J_2:$ dimensionless zonal harmonic coefficient of degree $\ell=2$ of the non spherically symmetric gravitational potential of the central body
\item[] $Q_2:$ dimensional mass quadrupole moment of the non spherically symmetric gravitational potential of the central body
\item[] ${\bds B}^\mathrm{oct}:$ gravitomagnetic spin octupole field in the empty space surrounding the rotating central body
\item[] $\phi^\mathrm{oct}:$ gravitomagnetic spin octupole potential in the empty space surrounding the rotating central body
\item[] ${\bds A}:$ perturbing acceleration experienced by the test particle
\item[] ${\bds r}:$ position vector of the test particle with respect to the central body
\item[] $r:$ distance of the test particle from the central body
\item[] ${\bds{\hat{r}}}:=  {\bds r}/r:$ radial unit vector
\item[] $\xi:=  \bds{\hat{k}}\bds\cdot\bds{\hat{r}}:$ cosine of the angle between the central body's spin axis and the position vector of the test particle
\item[] $\bds v:$ velocity vector of the test particle
\item[] $v_r:= \bds v\bds\cdot\bds{\hat{r}}:$ radial velocity of the test particle
\item[] $\lambda :=  \mathbf{\hat{k}}\mathbf{\cdot}\mathbf{v}:$ projection of the velocity of the test particle onto the direction of the spin axis of the central body
\item[] $\mathcal{P}_{\ell}\ton{\cdots}:$ Legendre polynomial of degree $\ell$
\item[] $a:$  semimajor axis of the test particle
\item[] $\nk :=  \sqrt{\mu/a^3}:$  Keplerian mean motion of the test particle
\item[] $\Pb:=  2\uppi/\nk:$ orbital period of the test particle
\item[] $e:$  eccentricity of the test particle
\item[] $p:=  a\ton{1-e^2}:$ semilatus rectum of the orbit of the test particle
\item[] $I:$  inclination of the orbital plane of the test particle to the plane $\grf{x,\,y}$ of some reference frame
\item[] $\Omega:$  longitude of the ascending node  of the test particle
\item[] $\omega:$  argument of pericenter  of the test particle
\item[] $\eta:$ mean anomaly at epoch
\item[] $f:$ true anomaly of the test particle
\item[] $u:=  \omega + f$ argument of latitude of the test particle
\item[] $\bds{\hat{l}}:= \grf{\cO,~\sO,~0}:$ unit vector directed along the line of the nodes toward the ascending node
\item[] $\bds{\hat{m}}:= \grf{-\cI\sO,~\cI\cO,~\sI}:$ unit vector directed transversely to the line of the nodes in the orbital plane
\item[] $\bds{\hat{h}}:= \grf{\sI\sO,~-\sI\cO,~ \cI}:$ normal unit vector such that $\bds{\hat{l}}\bds\times\bds{\hat{m}}=\bds{\hat{h}}$
\item[] $\bds{\hat{\uptau}}=\bds{\hat{h}}\bds\times\bds{\hat{r}}:$ transverse unit vector
\item[] $A_R:=  \bds A\bds\cdot\bds{\hat{r}}:$ radial component of the perturbing acceleration $\bds A$
\item[] $A_T:=  \bds A\bds\cdot\bds{\hat{\uptau}}:$ transverse component of the perturbing acceleration $\bds A$
\item[] $A_N:=  \bds A\bds\cdot\bds{\hat{h}}:$ normal component of the perturbing acceleration $\bds A$
\end{itemize}
\section{\textcolor{black}{Coefficients $\widehat{T}_j$ of the orbital rates of changes}}\lb{appenB}
\renewcommand{\theequation}{B\arabic{equation}}
\setcounter{equation}{0}
\textcolor{black}{
The coefficients $\widehat{T}_j,\,j=1,\,2,\,\dots 6$ entering \rfrs{dadtBS}{detdtBS} in Section\,\ref{AJ2} and \rfrs{dadtPS}{detdtPS} in Section\,\ref{Soct} are
}
\textcolor{black}{
\begin{align}
\widehat{T}_1 & := 1,\acap
\widehat{T}_2 & := \qua{\ton{\kl}^2 + \ton{\km}^2},\acap
\widehat{T}_3 & := \qua{\ton{\kl}^2 - \ton{\km}^2},\acap
\widehat{T}_4 & := \qua{\ton{\kh}\,\ton{\kl}},\acap
\widehat{T}_5 & := \qua{\ton{\kh}\,\ton{\km}},\acap
\widehat{T}_6 & := \qua{\ton{\kl}\,\ton{\km}}.
\end{align}
}
\end{appendices}
\bibliography{Megabib}{}
\end{document}